\documentclass[prb,superscriptaddress,floatfix,a4paper,aps,twocolumn,nofootinbib]{revtex4-1}

\usepackage{graphicx}
\usepackage{amsmath}
\usepackage{amssymb}

\usepackage{textcomp}

\usepackage{tabularx}
\usepackage{color}
\usepackage{xspace}

\usepackage[colorlinks=true,citecolor=blue,linkcolor=blue,urlcolor=blue]{hyperref}

\begin{document}

\title{Linear nonsaturating magnetoresistance in the Nowotny chimney ladder compound Ru$_2$Sn$_3$}

\author{Beilun Wu}
\affiliation{Laboratorio de Bajas Temperaturas y Altos Campos Magn\'eticos, Departamento de F\'isica de la Materia Condensada, Instituto Nicol\'as Cabrera and Condensed Matter Physics Center (IFIMAC), Unidad Asociada UAM-CSIC, Universidad Aut\'onoma de Madrid, E-28049 Madrid,
Spain}

\author{V\'ictor Barrena}
\affiliation{Laboratorio de Bajas Temperaturas y Altos Campos Magn\'eticos, Departamento de F\'isica de la Materia Condensada, Instituto Nicol\'as Cabrera and Condensed Matter Physics Center (IFIMAC), Unidad Asociada UAM-CSIC, Universidad Aut\'onoma de Madrid, E-28049 Madrid,
Spain}

\author{Federico Mompe\'an}
\affiliation{Instituto de Ciencia de Materiales de Madrid, Consejo Superior de
Investigaciones Cient\'{\i}ficas (ICMM-CSIC), Unidad Asociada UAM-CISC, Sor Juana In\'es de la Cruz 3,
28049 Madrid, Spain}

\author{Mar Garc{\'i}a-Hern{\'a}ndez}
\affiliation{Instituto de Ciencia de Materiales de Madrid, Consejo Superior de
Investigaciones Cient\'{\i}ficas (ICMM-CSIC), Unidad Asociada UAM-CISC, Sor Juana In\'es de la Cruz 3,
28049 Madrid, Spain}

\author{Hermann Suderow}
\affiliation{Laboratorio de Bajas Temperaturas y Altos Campos Magn\'eticos, Departamento de F\'isica de la Materia Condensada, Instituto Nicol\'as Cabrera and Condensed Matter Physics Center (IFIMAC), Unidad Asociada UAM-CSIC, Universidad Aut\'onoma de Madrid, E-28049 Madrid,
Spain}

\author{Isabel Guillam\'on*}
\affiliation{Laboratorio de Bajas Temperaturas y Altos Campos Magn\'eticos, Departamento de F\'isica de la Materia Condensada, Instituto Nicol\'as Cabrera and Condensed Matter Physics Center (IFIMAC), Unidad Asociada UAM-CSIC, Universidad Aut\'onoma de Madrid, E-28049 Madrid,
Spain}

\begin{abstract}We present magnetoresistivity measurements in high-quality single crystals of the Nowotny chimney ladder compound Ru$_2$Sn$_3$. We find a linear and nonsaturating magnetoresistance up to 20 T. The magnetoresistance changes with the magnetic field orientation at small magnetic fields, from a positive to a negative curvature. Above 5 T, the magnetoresistance shows no sign of saturation up to 20 T for any measured angle. The shape of the anisotropy in the magnetoresistance remains when increasing temperature and Kohler's rule is obeyed. We associate the linear and nonsaturating magnetoresistance to a small Fermi surface with hot spots, possibly formed as a consequence of the structural transition. We discuss the relevance of electron-electron interactions under magnetic fields and aspects of the topologically nontrivial properties expected in Ru$_2$Sn$_3$.\end{abstract}

\maketitle

\section{Introduction}

The transverse magnetoresistance (MR) shows the magnetic-field-induced changes in the electronic transport when the magnetic field is perpendicular to the current. It is defined as $MR=(\rho(B)-\rho(0))/\rho(0)$ and is usually small and positive. The magnetic field forces electrons into circular orbits and increases the resistivity, but usually by a small amount. Often, the MR presents a $B^2$ behavior, which is the simplest way to obey the Onsager relation $\rho(B)= \rho(-B)$. The cyclotron frequency $\omega_c=\frac{eB}{m}$ is proportional to magnetic field. At large values of $\omega_c\tau$ (with $\tau$ being the scattering time), increasing magnetic field essentially does not modify the mean-free path anymore and the MR saturates. When there is exact compensation between electron and hole-like carriers, this saturation does not occur and the MR continues as $B^2$[\onlinecite{Pippard1989}]. This peculiar situation is met in a few systems which have bottom and top of electron and hole bands very close to the Fermi level, such as WTe$_2$\cite{Ali2014}.  The observation of a linear MR at high magnetic fields in a single-crystalline metallic compound shows that electronic transport is influenced by features of the band structure which often produce noncircular paths on broken or peculiar orbits. For example, a linear MR can occur when the magnetic field is large enough to condense all electrons in the lowest Landau level, which occurs more easily in the presence of a linear dispersing bandstructure\cite{Abrikosov1998,Abrikosov2000,Kisslinger2015,Gopinadhan2015,Vasileva2016,Liang2014,Narayanan2015,Tang2011,Wang2012,Zhang2012,Assaf2013}. 
In some systems it might also occur at very specific magnetic field directions to allow for electron trajectories on open orbits\cite{wu2020huge}. 
A linear MR can also appear due to disorder in polycrystals of a system with open orbits or from a complex distribution of current paths across the sample under magnetic field\cite{Xu1997,Parish2003,Song2015}. A linear magnetoresistance is also consistently observed in many charge density wave systems or compounds with low electronic density and a small Fermi surface originated from folding at a structural or charge order transition \cite{Feng2019,Naito1982,Roetger1994,Luccas2015,Young2003,Budko1998,Galvis2013}.

\begin{figure*}[htbp]
\begin{center}
{\includegraphics[width = 1.8\columnwidth]{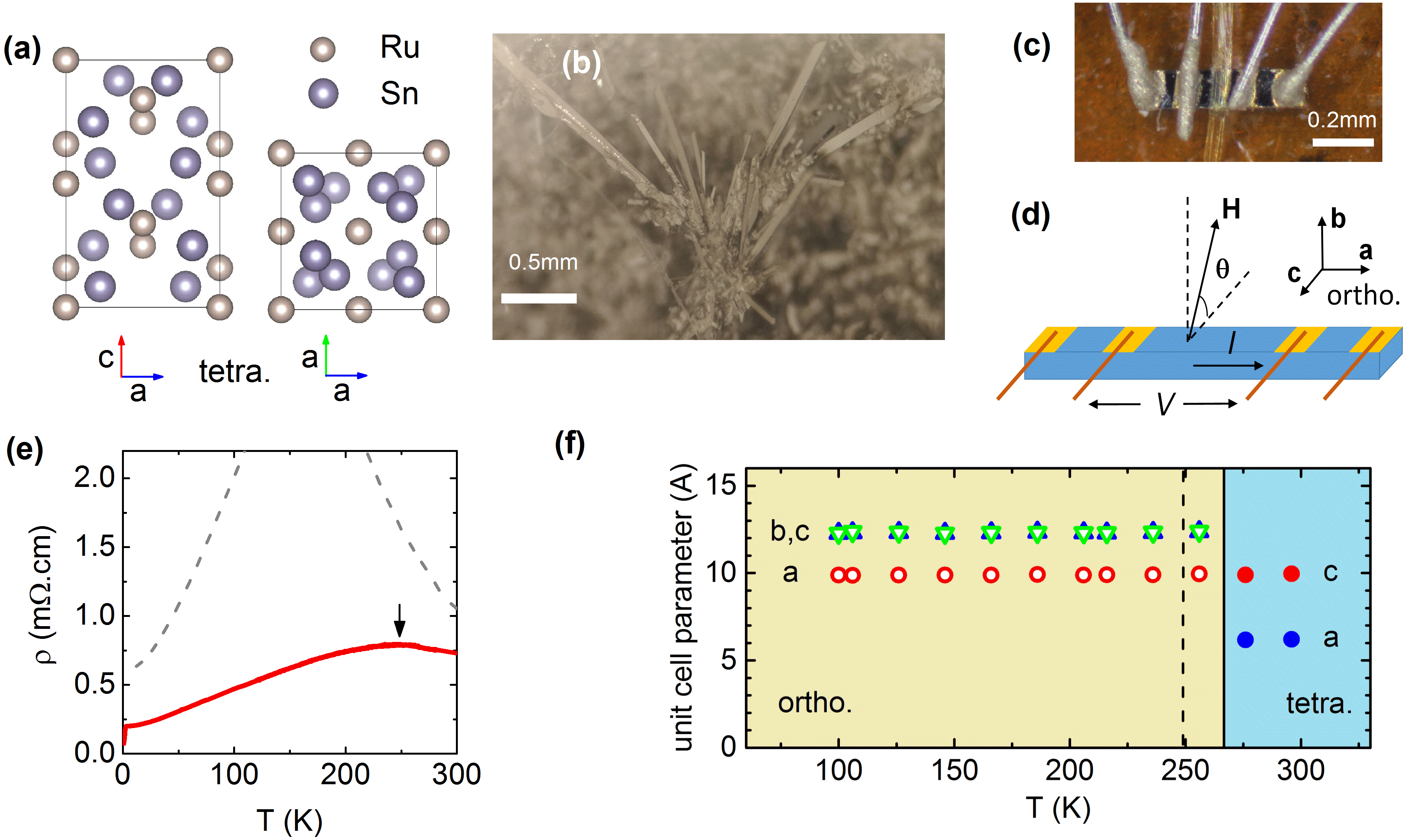}}
\caption{
\textbf{(a)} Crystal structure of the Ru$_2$Sn$_3$ in the high-temperature tetragonal phase. In the transition to the low-temperature orthorhombic phase, slight displacements of the Sn atom positions occur. The translational symmetry is lowered and the size of the unit cell doubled. Note the crystal axis notations are switched at the transition, following the correspondence $a \rightarrow b$ and $c$, $c \rightarrow a$.
\textbf{(b)} Photo of Ru$_2$Sn$_3$ crystals obtained after the growth. 
\textbf{(c)} One crystal with a needle shape has been isolated and contacted for transport measurements. We evaporated gold on the surface to improve contacts and used four silver paste contacts. A GE varnish filament is put on the middle of the crystal to provide mechanical support.
\textbf{(d)} We show in a scheme the direction of the applied current and magnetic field, together with the crystalline axis of the orthorhombic phase. The orientation of the crystal has been measured by single-crystal x-ray scattering.
\textbf{(e)} Resistivity vs temperature of a Ru$_2$Sn$_3$ single crystal. The temperature of the maximum in the resistivity is marked with an arrow. The dashed line shows the temperature dependence of the resistivity that we obtained in a sample where we did not correct for Sn deficiency after pre-melting (see text). This is close to the result of Ref.\citenum{Gibson2014,Shiomi2017}. 
\textbf{(f)} Cell parameters vs temperature of the single crystal. The broad transition from a tetragonal structure (blue region) to an orthorhombic structure (yellow region) is marked by the doubling of the lattice parameter $a$ of the tetragonal structure into the parameters $b$, $c$ of the orthorhombic structure. The dashed line corresponds to the temperature for the maximum in the $\rho (T)$ shown as an arrow in (e). 
}\label{FigureResistivity}
\end{center}
\end{figure*}

Ru$_2$Sn$_3$ belongs to the family of Nowotny chimney ladder (NCL) phases\cite{Poutcharovsky1975}. At room temperature it has a tetragonal crystal structure (Fig.\,\ref{FigureResistivity}(a)), and switches to an orthorhombic structure when cooled below room temperature\cite{Poutcharovsky1975,Susz1980}. It is among the few binary compounds where electron counting explains the chemical and structural stability of the compound\cite{Yannello}. Other examples include Ir$_3$Ga$_5$, RuGa$_2$, RuAl$_2$, Co$_2$Si$_3$ or Mn$_4$Si$_7$\cite{Fredrickson}. Most of the NCL systems are semiconductors, whereas Ru$_2$Sn$_3$ is a low-electronic-density metal\cite{Gibson2014,Susz1980}. 

In this study we report on angular dependent magnetoresistance measurements up to 20 T on Ru$_2$Sn$_3$ single crystals. We have grown single crystals of Ru$_2$Sn$_3$ and find a linear nonsaturating MR up to 20 T, which remains when increasing temperature up to 115 K. We also find an inversion of the MR curvature in the low-field region and discuss the relation between our observations and the electronic properties of this compound.

\section{EXPERIMENTAL}

We grew single-crystal Ru$_2$Sn$_3$ samples using a Bi flux method\cite{Canfield2010,Canfield2001,Canfield1992,canfield2019new}. To improve the mixing between the elements and due to the high melting temperature of Ru, we premelted the mixture containing Ru and Sn at stoichiometry (2:3) in an arc-melting furnace. We carefully measured the mass of the mixture precisely before and after premelting, and found a slight loss of mass. At the temperature of the arc furnace (2500$^\circ$C), the vapor pressure of Sn is more than six orders of magnitude higher than that of Ru. Therefore, the mass loss corresponds to the evaporation of Sn. We compensated the mass loss by adding Sn to the Ru-Sn mixture after premelting and mixed the result with Bi to act as a flux. The final mixture was set to a ratio of Ru:Sn:Bi=2:3:67\cite{Gibson2014}. We introduced it into an alumina crucible with a filter and a catch crucible (we used frit disk crucibles\cite{doi:10.1080/14786435.2015.1122248}) and then into an evacuated quartz ampoule with a residual Ar atmosphere. We first heated the growth to 1150$^\circ$C in 20 h, kept it at 1150$^\circ$C  for 42 h and then cooled it down slowly to 500$^\circ$C in 216 h. We then decanted the Bi flux in a centrifuge, and obtained bar-shaped crystals of Ru$_2$Sn$_3$ with a rectangular cross section and more than 1.5 mm length (Fig.\,\ref{FigureResistivity}(b)). 

We also made one growth without compensating for the Sn loss after premelting in the arc furnace. In this sample, we found similar results regarding the structural transition, the resistivity vs temperature and magnetoresistance (up to 9 T) as already reported in this material\cite{Kawasoko2014,Gibson2014}. In particular, the maximum in the resistivity was at about 170 K. In the sample where we have compensated for the Sn mass loss, the maximum in the resistivity occurs much closer to room temperature and is correlated with the structural transition measured with x-ray scattering on the same sample (in agreement with initial x-ray data in Ref.\onlinecite{Poutcharovsky1975}). Moreover, the broad maximum in the resistivity is strongly decreased and the residual resistivity is much smaller (Fig.\,\ref{FigureResistivity}(e)).

Thus the amount of Sn in the mixture is rather important in Ru$_2$Sn$_3$. In case of a concentration of Sn above stoichiometry, the remnant Sn is part of the flux and is removed when decanting. In the case of a small deficiency of Sn, this remains in the sample in the form of interstitial atoms or defects that might influence the structural transition and the resistivity.

In the following we focus on magnetoresistance experiments in the Sn-compensated sample. The sample is bar-shaped (Fig.\,\ref{FigureResistivity}(c)). The elongated direction is along the orthorhombic $a$ axis, and the side faces perpendicular to the $b$ and $c$ axis. The measurements have been made inside a cryostat capable of reaching 1 K, using methods described in Ref.\onlinecite{MONTOYA2019e00058}. The system is equipped with a 20 T (at liquid helium temperatures) superconducting magnet\cite{Magnet} and a rotating device to modify the angle of the applied magnetic field with respect to the sample\cite{wu2020huge}. For the resistivity measurements the usual four-probe AC method has been used, with the electrical contacts made by gluing fine gold wires with silver epoxy (Fig.\,\ref{FigureResistivity}(c)). In order to improve the contact resistance, four gold contacts have been Joule evaporated on the sample surface before gluing the gold wires.

\section{Results}

We have made single-crystal x-ray diffraction as a function of temperature. In Fig.\,\ref{FigureResistivity}(f) we show the obtained unit-cell parameters. Note that when the crystal structure changes from tetragonal to orthorhombic, we switch the notation of the crystal axes to $a \rightarrow b$ and $c$, $c\rightarrow a$. 
In our measurements, we did not resolve the difference between $b$ and $c$ axes in the orthorhombic structure. 
The structural transition is identified with the gradual appearance of new Bragg peaks corresponding to a doubling of the unit-cell parameters of the tetragonal structure. The intensity of these Bragg peaks increases gradually with decreasing temperature, as reported in Ref.\onlinecite{Poutcharovsky1975}.

Fig.\,\ref{FigureResistivity}(e) shows the temperature dependence of the resistivity $\rho (T)$. We find a maximum at $T=248$ K (marked by an arrow). This temperature is marked with a dashed line in Fig.\,\ref{FigureResistivity}(f) and is close to the structural transition temperature indicated by the appearance of the Bragg peaks of the low-temperature orthorhombic crystalline structure. We observe that the resistivity tends to decrease with temperature when cooling. The residual resistivity ($\rho_0=0.2$ m$\Omega$.cm) is much higher than in metallic systems and points out that Ru$_2$Sn$_3$ is a low-carrier-density semimetal, as we discuss below.

\begin{figure}[htbp]
\begin{center}
\includegraphics[width = 0.8\columnwidth]{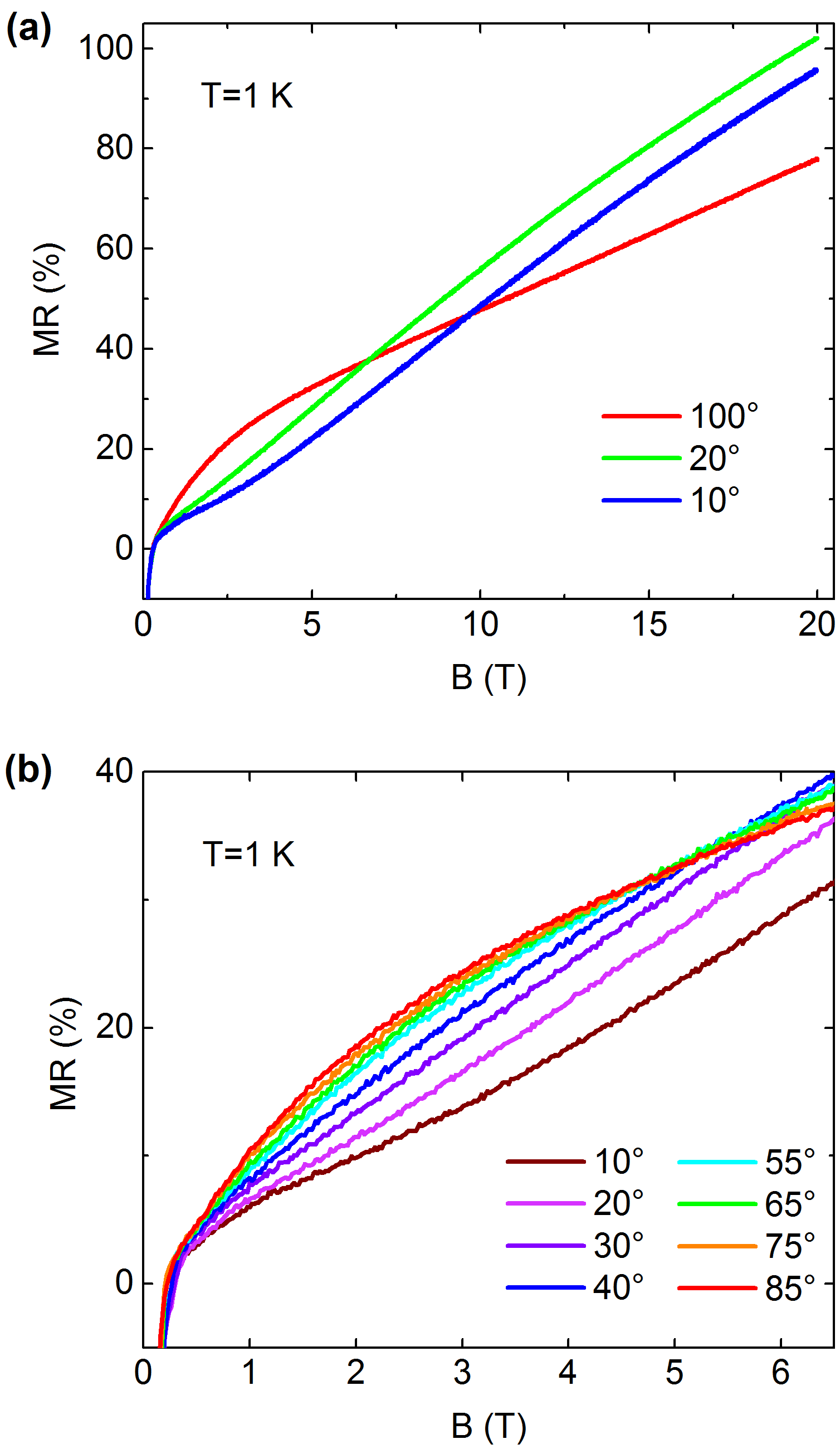}
\caption{\label{fig_MR_LT}
\textbf{(a)} We show (colored lines) the magnetoresistance  (MR $= (\rho(B)/\rho_0 -1) 100\%$) up to 20 T for different magnetic field orientations in the orthorhombic ($b$, $c$) plane (see Fig.\ref{FigureResistivity}(d)) at $T=1$ K. 
\textbf{(b)} Lines provide the MR in the low-field region, including data obtained for many more magnetic field orientations (at $T=1$ K). Notice the change of the curvature of the MR with the orientation.
}\label{FigureMagnetoresistance}
\end{center}
\end{figure}

\begin{figure*}[htbp]
\begin{center}
{\includegraphics[width = 2\columnwidth]{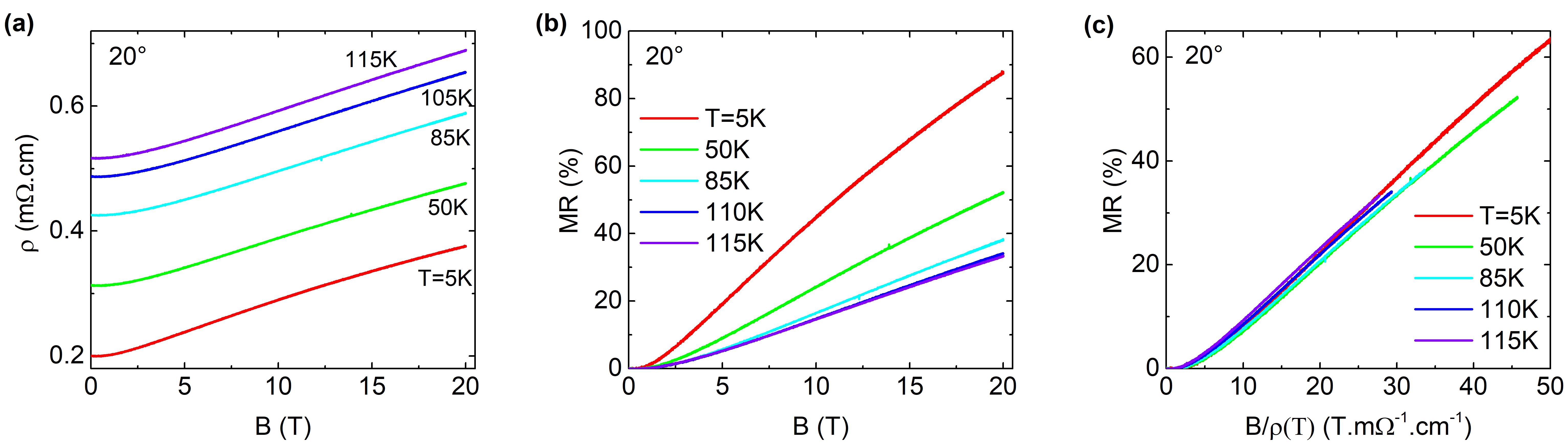}}
\caption{\label{fig_Tdep}
\textbf{(a)} We show as colored lines the resistivity vs magnetic field for different temperatures, with an orientation of the magnetic field corresponding to an angle $\theta = 20 ^{\circ}$ in the ($b$, $c$) plane. All data were taken in the low-temperature orthorhombic phase. 
\textbf{(b)} Same data, plotted as MR $= (\rho(B,T)/\rho(T) -1) 100\%$, with $\rho(T)$ the zero-field resistivity at the corresponding temperatures. \textbf{(c)} Same data, plotted in the form of a Kohler's plot. 
}\label{FigureKohler}
\end{center}
\end{figure*}

\begin{figure*}[htbp]
\begin{center}
{\includegraphics[width = 1.5\columnwidth]{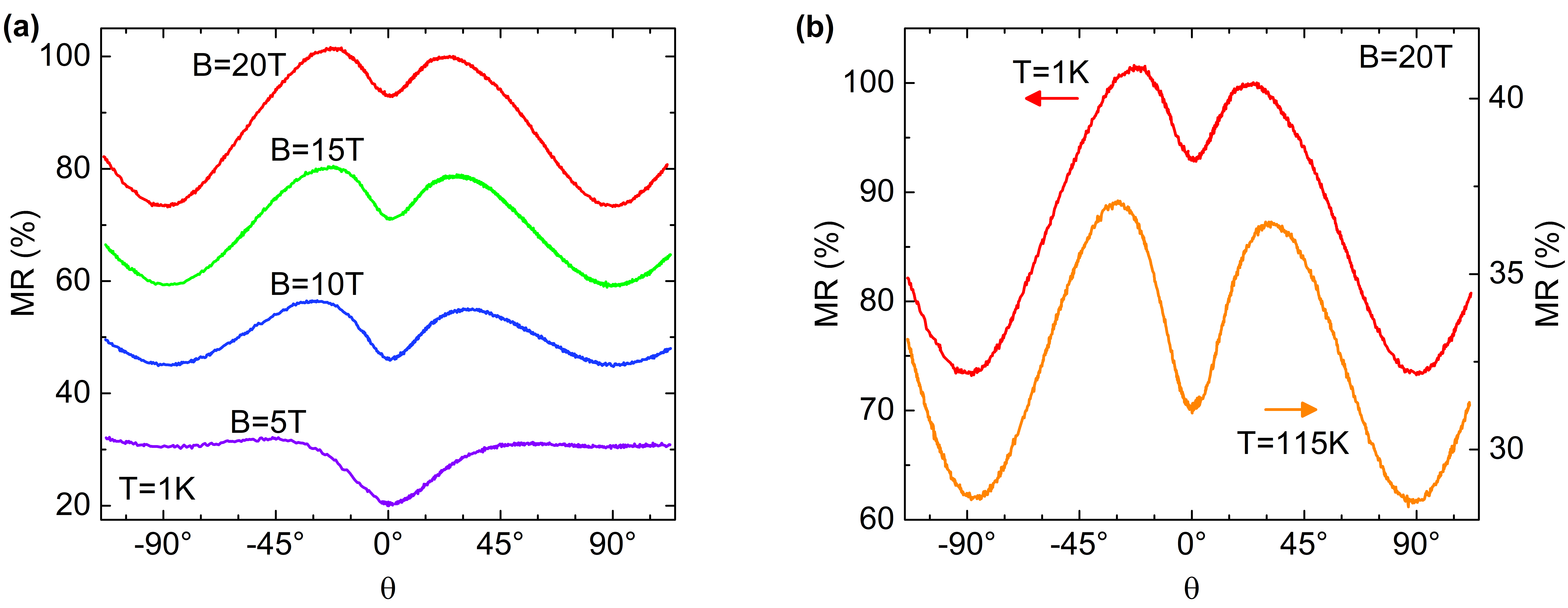}}
\caption{
\textbf{(a)} We show the angular dependence of the MR at different fixed magnetic fields and at $T=1$ K. \textbf{(b)} We show the angular dependence of the MR at $B = 20$ T for $T=1$ K and $T=115$ K.
}\label{fig_AngDep}
\end{center}
\end{figure*}

In Fig.\,\ref{FigureMagnetoresistance} we show the MR at $T=1$ K with different field orientations. The electrical current is along the orthorhombic $\textbf{a}$ axis, with the magnetic field oriented inside the ($b$, $c$) plane, always perpendicular to the current direction (Fig.\,\ref{FigureResistivity}(d)). In the field region below 0.2 T we see influence from the superconductinglike transition discussed in the Appendix. At 20 T the MR exceeds 100$\%$, much higher than in a previous magnetotransport study\cite{Shiomi2017}, and in our samples where the Sn mass loss was not compensated. This shows that the MR is, together with the temperature dependence of the resistivity, strongly dependent on the Sn content. The MR shows a linearlike behavior in the field region above 5 T at all the field orientations, while a clear inversion of the MR curvature can be seen at fields below 5 T. 

In Fig.\,\ref{FigureMagnetoresistance}(b) we show the field dependence of the MR in the low field region as a function of the field angle $\theta$. The MR evolves gradually from a quadratic-like behavior to a sublinear behavior with a downward curvature, when $\theta$ changes from 0$^{\circ}$ to 90$^{\circ}$. With this rotation, the field orientation (always perpendicular to the current) changes within the near-square ($b$, $c$) plane of the low-temperature orthorhombic phase. 

In Fig.\,\ref{FigureKohler}(a)-\ref{FigureKohler}(c) we show the magnetic field and temperature dependence of the MR for a fixed angle $\theta = 20 ^{\circ}$ well within the orthorhombic phase, up to $T=115$ K. Other angles show a similar behavior. Although the zero-field resistivity $\rho(T)$ increases by a factor of 3, the magnetic field dependence does not significantly change with the temperature. We find a linear MR up to the highest temperatures. We can see that the reduction in MR shown in Fig.\,\ref{FigureKohler}(b) scales with $B/\rho(T)$ (see Fig.\,\ref{FigureKohler}(c)). Thus Kohler's rule is obeyed in this temperature range. 

In Fig.\,\ref{fig_AngDep}(a) we show the MR as a function of the field angle in the ($b$, $c$) plane at different magnetic fields. We observe local minima at $\theta = 0 ^{\circ}$ and at $\theta = 90 ^{\circ}$. The MR has different values at the two minima, related to the anisotropy between the $b$ and $c$ axes in the orthorhombic phase. The depth of the minima changes with the magnetic field. In particular, the minima are best resolved  at $B=20$ T. In Fig.\,\ref{fig_AngDep}(b) we show the MR versus the field angle in the ($b$, $c$) plane at two different temperatures, at the same field $B=20$ T. We observe that the overall behavior of the angular dependence remains unchanged when increasing the temperature up to $T=115$ K, but the difference between the MR at the $\theta = 0 ^{\circ}$ minimum and the $\theta = 90 ^{\circ}$ minimum decreases. This suggests that the angular dependence is indeed related to the crystalline axes. With increasing temperature towards the structural orthorhombic-to-tetragonal transition, the orthorhombic anisotropy between the $b$ and $c$ axes is reduced.

\section{Discussion}

Let us start by discussing the temperature dependence of the resistivity presented in Fig.\,\ref{FigureResistivity}(e). In the low-temperature orthorhombic phase, $\rho (T)$ has a metallic behavior but shows a large residual resistivity of about 0.2 m$\Omega$cm. This value lies close to a heavily doped large-gap semiconductor or a doped narrow-gap semiconductor\cite{10.2307/2397982,Thompson1975,Sernelius1990}. Band structure calculations in both the orthorhombic phase\cite{Gibson2014,Kawasoko2014} and the high-temperature tetragonal phase\cite{Imai2005} show a wide gap at the Fermi level in the density of states. The gap, however, is not fully open, and there are a few bands crossing the Fermi energy. This feature is confirmed by the angle-resolved photoelectron spectroscopy (ARPES)\cite{Gibson2014} and optical conductivity\cite{Knyazev2018} measurements. The carrier density has been estimated to be of the order of $10^{20}$ cm$^{-3}$, at $T=2$ K from Hall coefficient measurements\cite{Shiomi2017}.

As we discuss above, previous studies\cite{Kawasoko2014,Shiomi2017,Gibson2014,Zhang2017,Susz1980,Fredrickson} suggest a strong dependence of $\rho (T)$ in Ru$_2$Sn$_3$ on the sample quality and composition. Notably,  Kawasoko \emph{et al.}\cite{Kawasoko2014} have shown that creating Sn deficiency in polycrystalline Ru$_2$Sn$_3$ samples leads both to a shift of the maximum in $\rho (T)$ to lower temperature and to an enhanced resistivity in the whole temperature range. Furthermore, thermopower measurements\cite{Gibson2014} show a remarkable change of the electron and hole carrier density with decreasing temperature, in samples with an enhanced resistivity. Applying hydrostatic pressure leads to a higher structural transition temperature and a smaller residual resistivity\cite{Zhang2017}.

 The structure of Ru$_2$Sn$_3$ presents interesting features that can be associated with this sensitivity of  Sn deficiency in the temperature dependence of the resistivity\cite{Poutcharovsky1975,Susz1980,Fredrickson}. In the room-temperature tetragonal phase, the 
Ru atoms form a fourfold helix. The unit cell has eight Ru atoms and the helix repeats each four Ru atoms. This follows the helical arrangement of the $\beta$-Sn structure. Sn atoms arrange in another helix embedded in the chimney formed by the Ru atoms and forming a one-dimensional structure with a threefold helix. Thus, there are two intertwined helices. The structural stability has been explained by taking the atomic arrangement as twinned layers formed by blocks with the RuGa$_2$-type structure, rotated at square angles to each other. The arrangement of blocks leaves too many Sn atoms in between layers. By removing those which have too high sterical pressure, the actual NCL structure is found\cite{Fredrickson}. Naturally, this arrangement leaves considerable room for incommensuration between the Ru and Sn sublattices and the nucleation of defects in the Sn structure\cite{Yannello,S.2000,E.2001}.

Now let us discuss the magnetic field dependence of the MR (Fig.\,\ref{FigureMagnetoresistance}). In the low field region below 5 T, and for field angles close to $\theta = 90 ^{\circ}$, we find a positive curvature with a near to quadratic increase of the MR, whereas for angles close to $0 ^{\circ}$ we find a negative curvature.

The most simple single-band MR expressions provide MR$\propto \frac{(\omega_c\tau)^2}{1+\beta(\omega_c\tau)^2}$ with $\beta$ a constant of order 1 \cite{Pippard1989}. This provides a quadratic MR for small $\omega_c\tau$ (small magnetic fields) which saturates at large $\omega_c\tau$ (large magnetic fields). In a two-band model with the presence of both electron and hole carriers, the MR is more complex, and it is found that the saturation at high magnetic fields can be lifted if the carrier number is exactly compensated, with a square nonsaturating field dependence\cite{Pippard1989,Ali2014}. Thus, the MR curvature observed at small magnetic fields might come from a modification of carrier concentration by the angle of the magnetic field, which leads to the MR increasing as $B^2$ for certain angles and saturating for others. The more remarkable result is, however, that neither the $B^2$ behavior nor the saturation continue for large magnetic fields. Instead, the magnetoresistance ceases to be quadratic or saturating and becomes linear for all field angles.

It has been shown that compounds with a charge density wave present a linear magnetoresistance already from very low magnetic fields due to strongly curved electron trajectories at pockets from small pockets originated by a Fermi-surface reconstruction due to the charge density wave\cite{Feng2019}. On the other hand, the behavior at high magnetic fields in the presence of a charge density wave can also be a linear MR. Hot spots on the Fermi surface, for instance at the ends of incomplete Fermi-surface contours formed by anisotropic charge density wave gap openings, lead to tunneling of electrons and trajectories which involve the vectors connecting the hot spots. The scattering rate is inversely proportional to the magnetic field at high fields and the MR shows a linear behavior\cite{PhysRevLett.29.124,PhysRevB.96.245129}. Considering that other NCLs are insulators at low temperatures, and that Ru$_2$Sn$_3$ is a low-carrier semimetal, it is unlikely that the Fermi surface in the low-temperature phase has enough small pockets to produce a complete gap opening, leading to linear magnetoresistance at small magnetic fields, as proposed in Ref.\onlinecite{Feng2019}. 

The MR behavior in Ru$_2$Sn$_3$ remains qualitatively when increasing temperature. As we see in Fig.\,\ref{FigureKohler}, Kohler's rule is obeyed. It states that in a metallic system, when no significant change occurs in the scattering processes which govern the dynamics of the charge carriers, the magnetic field $B$ induces a relative change of resistivity $\Delta \rho(B)/\rho(T)$ that scales with $B/\rho(T)$. In our case the Kohler's rule is verified up to temperatures above 100 K. This implies that the resistivity in this compound is not governed by phonon scattering.

It is useful to analyze the structural transition in more detail. The room temperature tetragonal structure of Ru$_2$Sn$_3$ has no inversion symmetry but transitions when cooling below room temperature into an orthorhombic structure with inversion symmetry\cite{Poutcharovsky1975,Susz1980}. The helix of Ru atoms (following the $\beta$-Sn structure) is left untouched at the transition, but Sn atoms show slight displacements when decreasing temperature. 
The orthorhombic $b$ and $c$ parameters become double that of the $a$ parameter of the tetragonal phase and the orthorhombic $a$ parameter becomes the $c$ parameter of the tetragonal phase. 
The same occurs in the semiconductors Ru$_2$Si$_3$ and Ru$_2$Ge$_3$, although at temperatures considerably above room temperature.  
The observed behavior is highly hysteretic at the transition, particularly in Ru$_2$Sn$_3$ over a temperature range of more than 300 K above room temperature\cite{Susz1980}. Nevertheless, below room temperature the system does not enter a metallic state immediately but shows a nonhysteretic broad maximum\cite{Kawasoko2014,Gibson2014}. 
As we discuss above, the verification of the Kohler's rule indicates that the change of phonon scatterings plays little role in $\rho$(T).  On the other hand, the doubling of the unit cell at the structural transition naturally leads to the folding of the Brillouin zone and a reconstruction of the Fermi surface. This might also be accompanied by a change of electron and hole charge carrier density, as suggested in the Hall effect and thermopower studies\cite{Kawasoko2014,Gibson2014}. The observed linear magnetoresistance in the low-temperature phase of Ru$_2$Sn$_3$ then might originate from a reconstructed Fermi surface at this structural transition, and the resulting enhancement of electron-electron scattering\cite{PhysRevLett.29.124,PhysRevB.96.245129}. 

On the other hand, Ru$_2$Sn$_3$ has been recently considered within the field of topological insulators\cite{Gibson2014,Shiomi2017}. The magnetoresistance is useful to investigate the electronic properties of topological insulators\cite{Hasan2010,Qi2011,Ando2013} and topological semimetals\cite{Ali2014,Liang2014,Shekhar2015}. In topological insulators, bulk and valence conduction bands are mostly due to bands derived from relatively isotropic atomic orbitals. Electronic band structure calculations in the low-temperature orthorhombic phase\cite{Gibson2014,Knyazev2018} reveal a small indirect gap between the Ru 4$d$ and Sn 5$p$ bands and suggest band inversion in the presence of spin-orbit coupling. ARPES studies observe a sixfold highly anisotropic Dirac state at the surface which has been associated to the directional orbital character of the Ru- and Sn-derived inverted bands\cite{Gibson2014}. This is in contrast to the Bi-based three dimentional topological insulators\cite{Hasan2010,Qi2011,Ando2013}, where a $s$-$p$ band inversion opens a gap in the bulk state leaving a surface edge state with a two-dimentional Dirac cone\cite{Hsieh2009,Xia2009}. A possible explanation for a linear MR here is then related to these Dirac surface states, as in the Bi-based topological insulators\cite{Tang2011,Wang2012,Zhang2012,Assaf2013}. However, it is difficult to isolate the contribution of surface states to the resistivity in Ru$_2$Sn$_3$. At present, it remains unclear how these anisotropic behaviors influence the magnetoresistance. 
The angular dependent magnetoresistance does not show a strong anisotropy as expected for a two-dimensional surface transport channel, but instead reveals an orthorhombic symmetry pattern of the bulk states.

\section{Conclusions}

In summary, we have grown single crystals of Ru$_2$Sn$_3$ and performed angular-dependent magnetotransport measurements up to 20 T. We find a strong dependence of the transport properties on the Sn content in the initial composition of Ru$_2$Sn$_3$, indicating that the role of defects and Sn vacancies is rather important in understanding the properties of this compound. We find a linearlike and nonsaturating MR, of larger magnitude than previous works, which persists up to above 100 K. The MR curvature evolves strongly with the field angle at low fields. We associate the linear MR to strong electron-electron scattering due to a small-sized Fermi surface in the low-temperature crystalline phase.

\section*{Acknowledgement}

This work was supported by the Spanish Research State Agency (FIS2017-84330-R, CEX2018-000805-M, RYC-2014-15093, MAT2017-87134-C2-2-R), by the Comunidad de Madrid through the program NANOMAGCOST-CM (Grant No. S2018/NMT-4321) and by the European Research Council PNICTEYES project through Grant Agreement No. 679080. We acknowledge SEGAINVEX at UAM for the design and construction of the cryogenic equipment and support from COST Action CA16218 (Nanocohybri). We also thank R. \'Alvarez Montoya and J. M. Castilla for technical support and S. Delgado for support in the single-crystal synthesis. We acknowledge SIDI at UAM for support with regard to sample characterization. 
\bigskip
\bigskip

\section*{Appendix}

\begin{figure}[htbp!]
\begin{center}
{\includegraphics[width =0.8\columnwidth]{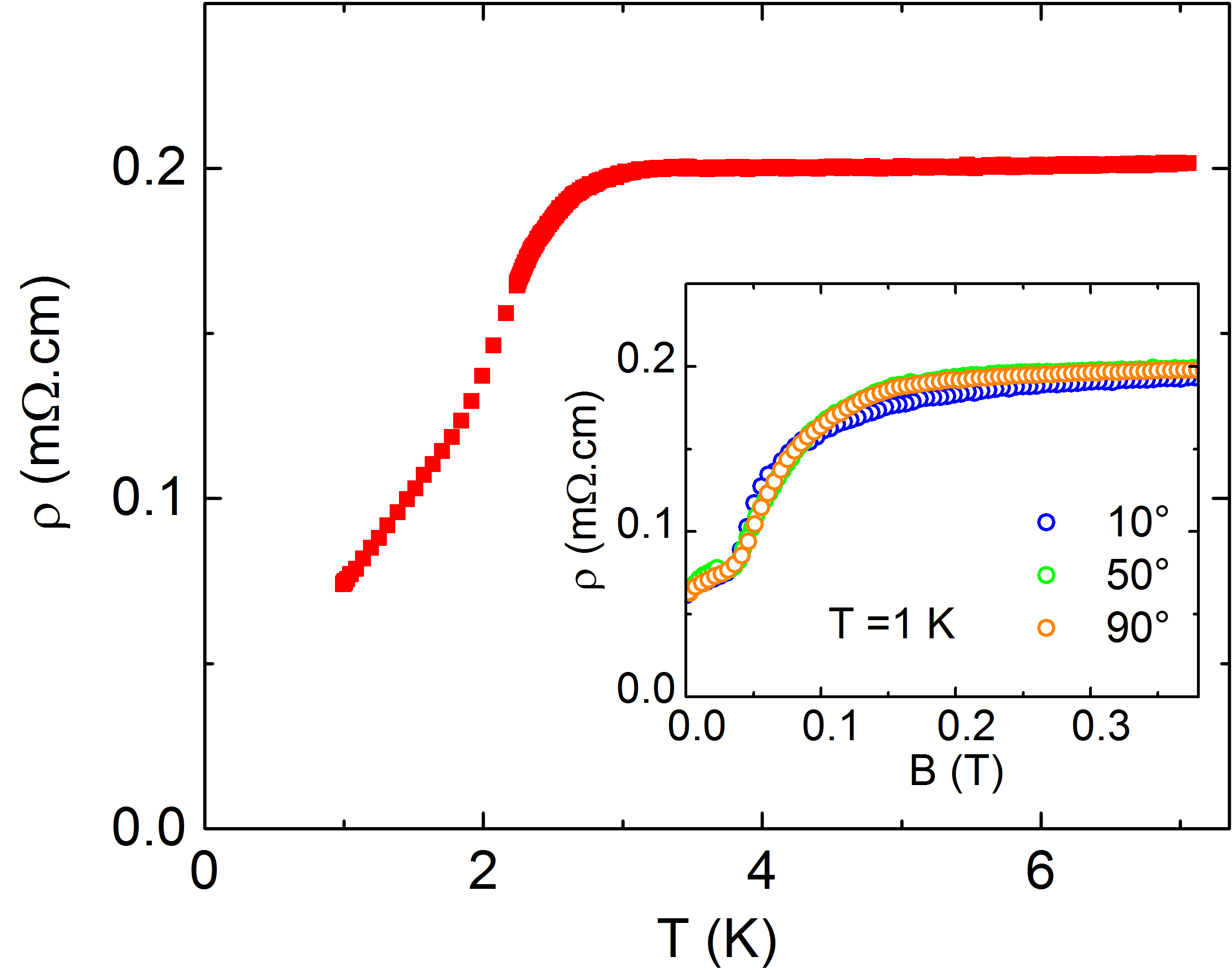}}
\caption{In the main panel we show the resistivity vs temperature as red filled squares. We notice the onset of an incomplete superconducting transition. In the inset we show the magnetic field dependence of the resistivity for small magnetic fields as colored circles. The magnetic field is applied with the angle $\theta$ shown in the legend. See also Fig.\,\ref{FigureResistivity}(e). 
}\label{FigureSuperconduct}
\end{center}
\end{figure}

We deem it relevant to note that at temperatures below 3 K we observe a strong decrease in the resistivity, which suggests an incomplete superconducting transition. $\rho(T)$ does not become zero at about 1 K (Fig.\,\ref{FigureSuperconduct}). In the inset of Fig.\,\ref{FigureSuperconduct} we show that this superconducting-like transition has a critical field of the order of 0.1 T and presents little angular dependence when the field is rotated in the orthorhombic ($b$, $c$) plane. One possibility is that this transition comes from traces of Sn residues, as observed and discussed in Ref.\onlinecite{Shiomi2017}. However, the onset of the decrease in $\rho(T)$ occurs here well below 3 K, whereas the superconducting transition temperature of Sn is of $T_c = 3.75$ K. Other options are that there are traces of oxides or small inclusions of an unknown Sn-based mixture, or that Ru$_2$Sn$_3$ itself might show superconductivity if the crystal quality is further improved.
\bigskip

\bibliographystyle{apsrev4-1-titles}

%

\end{document}